\documentstyle[prl,aps,epsf,twocolumn]{revtex}
\newcommand{\eq}{\begin{equation}}
\newcommand{\ee}{\end{equation}}

\newcommand{\vrr}{{\vec{r}}}

\newcommand{\vq}{{\vec{q}}}
\newcommand{\vQ}{{\vec{Q}}}

\def\lam{\lambda}
\def\dd{d^{\dagger}}

\def\half{{1\over2}}
\def\third{{1\over3}}
\def\twof{{2\over5}}
\def\threes{{3\over7}}
\def\rhob{{\bar \rho}}
\def\ua{\uparrow}
\def\da{\downarrow}
\def\eqa{\begin{eqnarray}}
\def\eea{\end{eqnarray}}
\begin{document}
\draft
\flushbottom
\twocolumn[
\hsize\textwidth\columnwidth\hsize\csname @twocolumnfalse\endcsname
\title{ Hamiltonian Description of Composite Fermions: Calculation of
Gaps }
\author{  Ganpathy Murthy$^{a}$ and  R.Shankar$^{b}$  }
\address{
{\it (a)} Physics Department, Boston University, Boston MA 02215\\
and
Department of Physics and Astronomy, Johns Hopkins University,
Baltimore MD 21218;\\ {\it (b)} Department of Physics, Yale
University, New Haven CT 06520}
\date{\today}
\maketitle
\tightenlines
\widetext
\advance\leftskip by 57pt
\advance\rightskip by 57pt

\begin{abstract}
 We analytically calculate gaps for the $\nu=\third, \
\twof$, and the $\threes$ polarized and partially polarized  
Fractional
Quantum Hall (FQH) states based on the Hamiltonian Chern-Simons  
theory
we have developed. For a class of potentials that are soft at high
momenta (due to the finite thickness of the sample) we find good
agreement with numerical and experimental results.
\end{abstract}
\vskip 1cm

]

\narrowtext
\tightenlines
The Composite Fermion(CF) picture\cite{jain-cf} has been a very
successful organizing principle in understanding the Fractional
Quantum Hall effects(FQHE)\cite{fqhe-ex} as well as in generating
wavefunctions
for these states\cite{jain-cf-review}.  Chern-Simons(CS) field
theories\cite{gcs,gmcs,zhk,read1,lopez} have provided us with a link
between
the microscopic formulation of the problem and experiment, both for
incompressible and compressible states\cite{kalmeyer,hlr}.

Recently we developed a hamiltonian CS theory for the FQH states
\cite{us1,us2}. Inspired by  the work
of Bohm and Pines\cite{bohm-pines} on the 3D electron gas, we
enlarged
the Hilbert space to introduce $n$ high-energy magnetoplasmons
degrees
of freedom, ($n$ also being the number of electrons) at the same time
imposing an equal number of constraints on physical states.  Upon
ignoring the coupling between the oscillators and the fermions we
obtained some well known wavefunctions
\cite{laugh,jain-cf}. However the fermions still had the bare mass,
and the frequency of the magnetoplasmons was incorrect. Hence a final
canonical transformation was employed to decouple the fermions from
the oscillators {\em in the infrared limit}.

	We choose to call the final fermions {\em the} composite
fermions for the following reasons. First, the final canonical
transformation assigns to each fermion the magnetic moment $e/2m$ as
mandated by the arguments of Refs.
\cite{ssh,simon-f1}.  Next, the fermions have $1/m^*=0$ in the
absence of interactions and acquire an interaction dependent
$1/m^*$. Finally, the formula for the electronic charge density takes
the following form, separable into high- and low-energy
pieces\cite{us2}, at small $q$:
\eqa
\rho_e(q)={q \over \sqrt{8\pi}} \sqrt{{2p\over2p+1}} (A(q)+
A^{\dagger}
(-q))&\nonumber\\
+{\sum_j e^{-iqx_j} \over 2p+1} -{il_{}^{2} }  (\sum_j (q \times
\Pi_j)e^{-iqx_j}&
)\label{rhobar}
\eea
where $A$ refers to the oscillators, $l =1/\sqrt{eB}$ is the magnetic
length, and ${\vec\Pi}_j={\vec P}_j+e{\vec A}^*(r_j)$ is the velocity
operator of the CFs.  The oscillator piece saturates Kohn's
theorem\cite{kohn}, and the rest, to be called $\rhob$, satisfies the
magnetic translation algebra\cite{GMP} to lowest leading order. Note
that $\rhob$ is a sum of a monopole with charge $e^*=e/(2p+1)$, which
is the charge associated with the CF, and a dipole piece which alone
survives at $\nu=1/2$ and has the value proposed by Read\cite{read2}.
(A number of recent constructions have emphasized this dipolar
aspect\cite{dh,pasquier}).

 The Hamiltonian of the low-energy sector  (dropping the magnetic
moment term)
is
\eq
H=\half \int {d^2 q\over(2\pi)^2} v(q) \rhob(-q) \rhob(q)\label{ham}.
\ee
where $v(q)$ is the electron-electron interaction.
This hamiltonian is to be supplemented by $n$ constraints (which will
be ignored in the earlier parts of this discussion).

Earlier, we extracted \cite{us1,us2} from this Hamiltonian a
``free-particle'' part, (corresponding to the diagonal terms in the
sum over particles ) and thereby an effective mass for the CFs, which
we found (for a Coulomb interaction cutoff at wave vector
$Q=k_F=1/l $) to be $ {1\over m^*}\simeq {e^2\over \epsilon l }
\cdot
{1\over 6} $ where $\epsilon$ is the dielectric constant. However, we
emphasized\cite{us2} that since the interactions were of the same
magnitude as the ``free-particle'' term and could renormalize this
mass considerably, one should calculate directly observable  
quantities
using the full Hamiltonian.  This is what we shall do now by  
computing
transport gaps for a few fractions. Since the CF already has many of  
the
nonperturbative charge and mass renormalizations built into it, we  
can
expect that we may pass readily from the hamiltonian to observables
using a variety of approximation methods.

We now describe our calculation of gaps for $\nu =\third,\ \twof,\
\threes$ in the Hartree-Fock (HF) approximation. The gaps are given  
by
the difference in the expectation of $H$ between the ground state and
a state with a widely separated particle-hole pair.  For the states
that appear in the calculation, we use the ones that appear naturally
in CF theory\cite{jain-cf}, which are just the eigenstates of the
free-particle part alluded to earlier: $p$ filled CF- Landau levels
(CF-LLs) for the ground state, and states with particle-hole
excitations above it.

Note that we do not convert these CF wavefunctions to electronic ones
(by multiplying by Jastrow factors and projecting to the LLL).
Instead we use our formula for the electronic charge operator
$\bar{\rho}$ (in which these effects are subsumed by the reduced
charge $e^*$ and the dropping of the oscillator part) and write $H$  
in
terms of it.  We circumvent the fact that $\bar{\rho}$ is to be
trusted only for $ql \ll 1$ as follows.  Consider real samples which
have a finite thickness $\Lambda$ of the same order as $l $, so that  
the
Coulomb interaction is cutoff at large $q$\cite{thick1}. It was
realized by Haldane and Rezayi\cite{hal-rez} that this has a large
effect on the gap, while leaving the wavefunctions essentially
unchanged. We will focus on such interactions, parametrized by $\lam
=\Lambda /l $ for which numerical results are
available\cite{thick2,bonesteel,jain-th}. The advantage is that as
$\lam$ becomes large only small-$q$ matrix elements of the density  
are
invoked in computing gaps, and we expect our theory to become more
accurate. It is possible that beyond some large $\lam$ the liquid
state might cease to be the ground state. Our theory, which is based
on a liquid state with uniform density, can be expected to work up to
this $\lam$.

 We work in the symmetric gauge, in which the single-particle
wavefunctions are characterized by the LL index $n$ and the angular
momentum index $m$.  The formalism in this gauge has been extensively
developed\cite{laugh,fetter}. The magnetoexciton wave
functions\cite{lerner} $\psi_{n}^{n'}(q;\vrr,\vrr')$ describe a
particle in the $n^{th}$ LL and a hole in the $n'^{th}$ LL, with a
conserved momentum $\vq$. In a second quantized notation we can
introduce creation ($\dd_{nm}$) and destruction operators ($d_{nm}$),
and write the magnetoexciton creation operator as $
X_{n}^{n'}(q)= \sum\limits_{mm'} x_{nn'}^{mm'}(q)
\dd_{nm}
d_{n'm'} $ where the $x$-coefficients are (for $m\ge m'$)

\eqa
x_{nn'}^{mm'}(q)=(-1)^{n'}{\sqrt{2\pi}\over L}\sqrt{{m'!\over
m!}}&\bigl({iQ_+\over \sqrt{2}}\bigr)^{m-m'} \nonumber \\
\times L_{m'}^{m-m'}(y)e^{-y/2}&
\eea
Here $Q_+=l^*(q_x+iq_y)$, $l^*=l \sqrt{2p+1}$ is the magnetic length
in the effective field, $L_n^m$ is a Laguerre polynomial, and
$y=Q^2/2$.  The density operator can now be written in terms of the
above operators as
\eqa
\rhob(q)=\sum\limits_{nn'} \rho_{n}^{n'}(q)X_{n}^{n'}(q)&\nonumber\\
\rho_{n}^{n'}(q)={(-1)^{n'+1}\over
2p+1}{L\over\sqrt{2\pi}}&\sqrt{n'!\over
n!}\bigl({-iQ_+\over\sqrt{2}}\bigr)^{n-n'}\nonumber \\
\times
e^{-y/2}(nL_{n'-1}^{n-n'}+2L_{n'}^{n-n'}&-(n'+1)L_{n'+1}^{n-n'})
\eea
Apart from the trivial dependence of $x$ on $n'$, we see the
separation between
the angular $m$ labels and the
``radial'' $n$ labels. {\em We find that  that as $\vq\to 0$ all
transition matrix elements vanish at least as $q^2$, an essential
property of charge density matrix elements in the LLL\cite{GMP}.}

We will  need the following  identities;
\eqa
\sum\limits_{m}^{}x_{nn}^{mm}(\vq) =(-1)^n
&{L\over\sqrt{2\pi}}&\delta_{{\vec Q},0}\nonumber\\
\sum\limits_{m}^{}
x_{n_1n}^{m_1m}(\vq_1)x_{nn_2}^{mm_2}(\vq_2)&=(-1)^n&{\sqrt{2\pi}\over 
 L}e^{{-i\over2}\vQ_1\times{\vQ}_2}\times\nonumber\\
&&x_{n_1n_2}^{m_1m_2}(\vq_1+\vq_2)\label{identities}
\eea
The ground state we have assumed is not an eigenstate of $H$, which
can create particle-hole pairs above it. However, for rotationally
invariant interactions, we find that the Hamiltonian does not mix a
single-particle (or single-hole) state with any other single-particle
(or single-hole) state, the signature of a HF state. This supports  
our
view that we are working with the right variables.

Consider  transport gaps
in the spin-polarized states of $\third$, $\twof$, and $\threes$.  We
will work with the model potential $v(r)=e^2/\varepsilon\sqrt{r^2+\Lambda^2}$,  
whose
Fourier transform is $v(q)=2\pi e^2
\exp{(-q \Lambda
   )}/\varepsilon q$ where $\Lambda$ is the thickness.  In order to compute the
gap
to a neutral excitation (with widely separated quasiparticle and
quasihole), which corresponds experimentally to the transport gap, we
compute the energy to add a single CF to the $p^{th}$ LL, and add to
it the energy to remove a CF from the $(p-1)^{th}$ LL. To illustrate
how the calculation goes, we work out the case of the quasihole
energy
for $\nu=\third$. The quasihole state is $d_{0m}|\Omega>$, where
$|\Omega>$ denotes the GS. Now one has to compute

\begin{eqnarray}
&\sum\limits_{n_in_i',m_im_i'} x_{n_1n_1'}^{m_1m_1'}(-q)
x_{n_2n_2'}^{m_2m_2'}(q) \rho_{n_1}^{n_1'}(-q)
\rho_{n_2}^{n_2'}(q)\nonumber\\
&<\Omega|\dd_{0m} \dd_{n_1m_1} d_{n_1'm_1'}\dd_{n_2m_2}d_{n_2'm_2'}
d_{m0}|\Omega>_c
\end{eqnarray}
where the subscript $c$ on the expectation value denotes that the
operators of the external particle are not allowed to contract with
themselves (this corresponds to subtracting the GS energy).  Now one
performs the standard Wick contractions, with the proviso that two
operators belonging to the same density are not allowed to contract
with each other (this would correspond to the $q=0$ density, which is
cancelled by the background). After performing the Wick contractions
one is left with sums over the angular momentum indices, which can be
performed by using Eqs.(\ref{identities}).

Similar considerations hold for the quasiparticle energy. Finally,
adding the two  together to obtain the gap, leads to the following
expressions
\begin{eqnarray}
\Delta(\third)=&K_3\bigl({l^{*3}\over
\Lambda^3}+h_0+2h_1+h_2-h_3\bigr)\nonumber \\
\Delta(\twof)=&K_5\bigl({l^{*3}\over
\Lambda^3}+h_0+4h_1+{3\over2}h_2-10h_3+{9\over2}h_4-\half
h_5\bigr)\nonumber \\
\Delta(\threes)=&K_7\bigl({l^{*3}\over
\Lambda^3}+h_0+6h_1+h_2-{104\over3}h_3+{
425\over12}h_4-{25\over2}h_5\nonumber \\
+&{7\over4}h_6-{1\over12}h_7\bigr)\nonumber\\
&h_n={1\over K_p(2p+1)^2} \int {d^2 q\over(2\pi)^2} v(q) y^n
e^{-y}\label{deltas}
\end{eqnarray}
 where $K_p=e^2/(\varepsilon l (2p+1)^{5/2})$, and $y=q^2l^{*2}/2$. We
can compute the gaps for {\it any} potential that has its third  
moment
finite\cite{fang-howard} (this leads to the first term in the gaps).

 Recently, numerical calculations based on CF wavefunctions have been
performed\cite{bonesteel,jain-th} on the model potential. Since these
wave functions have essentially perfect overlap with exact ground
states\cite{jain-cf-review}, we surmise that the results are
essentially exact. Bearing in mind that our results are to be trusted
for large $\lam=\Lambda /l $, we find for all three fractions an  
error of
about 60\% at $\lam=2$ which decreases to about 30\% for $\lam=3$.

Table I compares our numbers to those measured on an experimental
sample\cite{du-numbers}, whose $\lam$ has been determined as  
explained
in ref.\cite{jain-th}, based on the detailed calculations of
ref.\cite{ortalano}. The difference ($\simeq 10-30\% $) between
Ref.\cite{jain-th} and experiment is presumably due to disorder,
Landau-level mixing etc.
\vskip 0.15in
\begin{tabular}{|c|c|c|c|c|c|}
\hline
 $\nu$ &$\lam$ & Ref.\cite{du-numbers} & Ref.\cite{jain-th}& Ours1&  
Ours2
\\
\hline
 1/3 & 2.5 & .042 & .046 & .066 & .053  \\
\hline
 2/5 & 2.3 & .017 & .021 & .031 & .020  \\
\hline
 3/7 & 2.2 & .010 & .015 & .020 & .011  \\
\hline
\end{tabular}

\narrowtext
\vskip 0.15in
Table I: Gaps in units of $e^2/\varepsilon l$. The comparison is
between experiment (Ref.\cite{du-numbers}), numerical work
(Ref.\cite{jain-th}), and the results of the preceding
calculations. The last column corresponds to our results with
constraints, to be explained later. \\

We have also computed the gap to the spin-reversed quasiparticle for
$\nu=\third$ (ignoring the Zeeman energy),
$\Delta_{SR}=K_3(h_0+2h_1+h_2)$. The only results we are aware of for
the model potential(ref(\cite{bonesteel}) give values for this
quantity for $\lam=0, 1.5$.  The error at $1.5$ is about 30\%.

Apart from numbers for the gaps, we have also derived scaling
relations between gaps for $p/(2p+1)$ and $p/(2sp+1)$, which we will
present in detail elsewhere\cite{kwon-us}.
\begin{figure}
\narrowtext
\epsfxsize=2.4in\epsfysize=2.4in
\hskip 0.3in\epsfbox{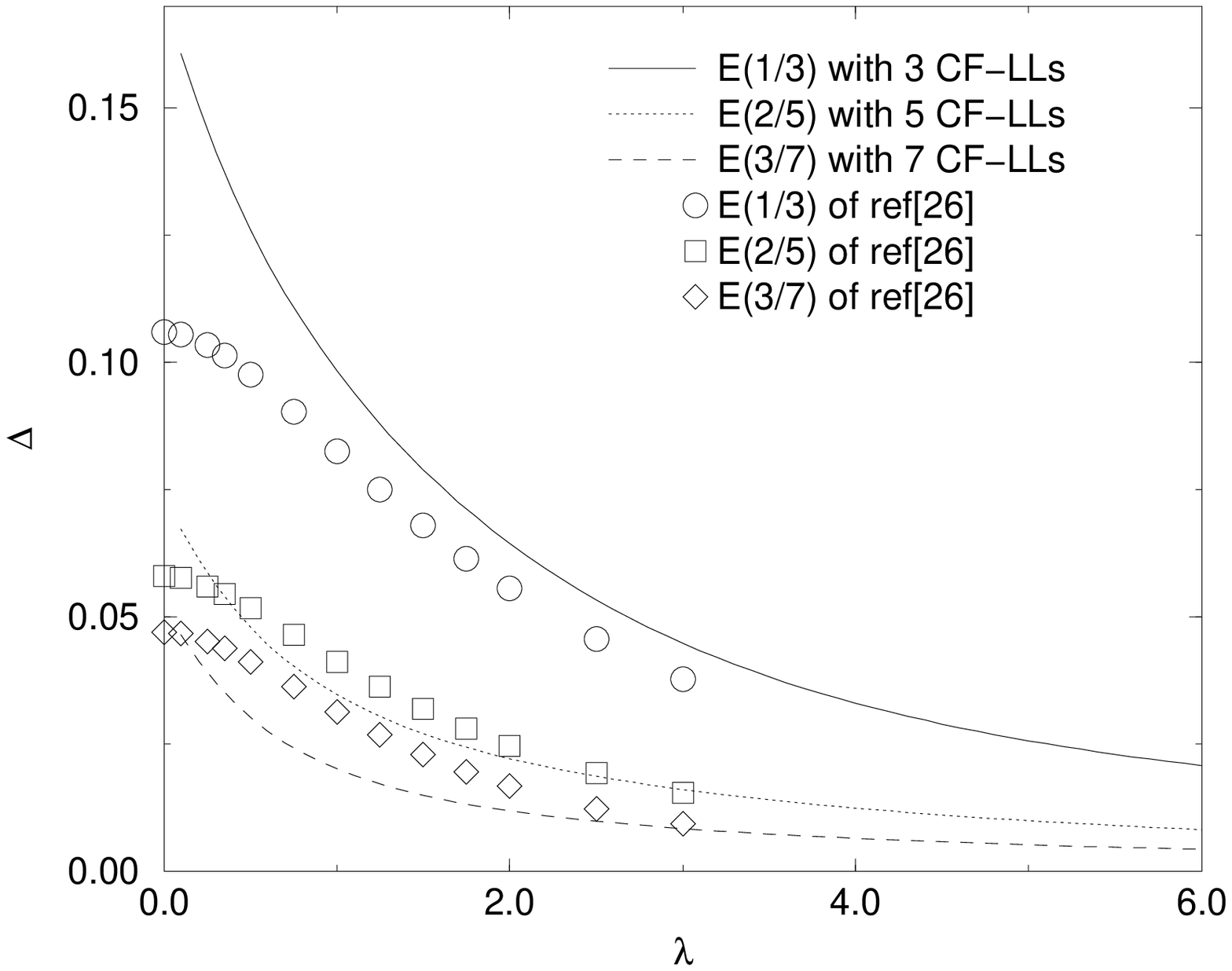}
\vskip 0.15in
\caption{Gaps to polarized $\third$, $\twof$, and $\threes$ states
and
comparison to results of ref.[26]. All energies are in units of
$e^2/\varepsilon l$.
\label{fig1}}
\end{figure}

 Now let us consider the constraints.  The divergence of gaps at
small
$\lam$ comes from summing over all CF-LLs in the sum over
intermediate
states.  We know that the Hilbert space of free fermions is too
large for this problem, which is restricted to the electronic LLL. It
is the role of the constraints, ignored so far, to select the
suitable
subspace.  In an ideal calculation,  a gauge invariant charge
operator
acting on a physical gauge invariant state will not mix it with
intermediate states which are unphysical. There will be no need to
invoke constraints.  Given that our $\bar{\rho}$ obeys the magnetic
algebra when the commutators are calculated in the big space with no
regard to constraints, we may conclude it is gauge invariant.
However,
there are two problems: It represents the density only for small $q$,
and the trial states we sandwich it between are free
particle, non-gauge invariant, states.

While we do not know which states will be selected by the
constraints,
we know that (i) in numerical work \cite{dev} the low-lying states
are
in one-to-one correspondence with those of independent CFs, and (ii)
the number of constraints is so as to limit the single particle
states
to those of one filled electronic LL or $2p+1$ filled CF levels.
This
suggests a simple, approximate way of dealing with the constraints:
keep the first $2p +1$ lowest-lying CF-LL states. Our results, as
shown in Fig.1 in this approximation, are in reasonable agreement
with
those of ref.\cite{jain-th} over a whole range of $\lam$.  The last
column of Table I displays the numbers for the appropriate $\lam$.  
While the
agreement is gratifying, this is not a controlled procedure, and we
cannot be sure that other quantities computed this way will come out
with the same accuracy.

We have also computed gaps to singlet (for $\twof$) and partially
polarized states (for $\threes$), ignoring the Zeeman energy, and
imposing the constraint as described above. We choose the many-body  
GS
for the singlet $\twof$ state in accordance with CF
theory\cite{jain-cf} to be the $n=0$ CF-LL occupied with both $\ua$
and $\da$ spin CFs. There is only a single energy gap in this system,
since both the particle and hole energies are symmetric in spin. For
partially polarized $\threes$ we choose the GS to be the $n=0$ CF-LL
occupied by both spins, while the $n=1$ CF-LL is occupied by $\ua$
spins\cite{jain-cf}. In the latter case there are distinct energies
for the four possible excitations, with the particle being $\ua$
($n=2$ CF-LL), or $\da$ ($n=1$ CF-LL), and the hole being $\ua$  
($n=1$
CF-LL) or $\da$ ($n=0$ CF-LL). Our results are presented in Fig.2.   
We
are not aware of calculations of these quantities for the model
potential.

\begin{figure}
\narrowtext
\epsfxsize=2.4in\epsfysize=2.4in
\hskip 0.3in\epsfbox{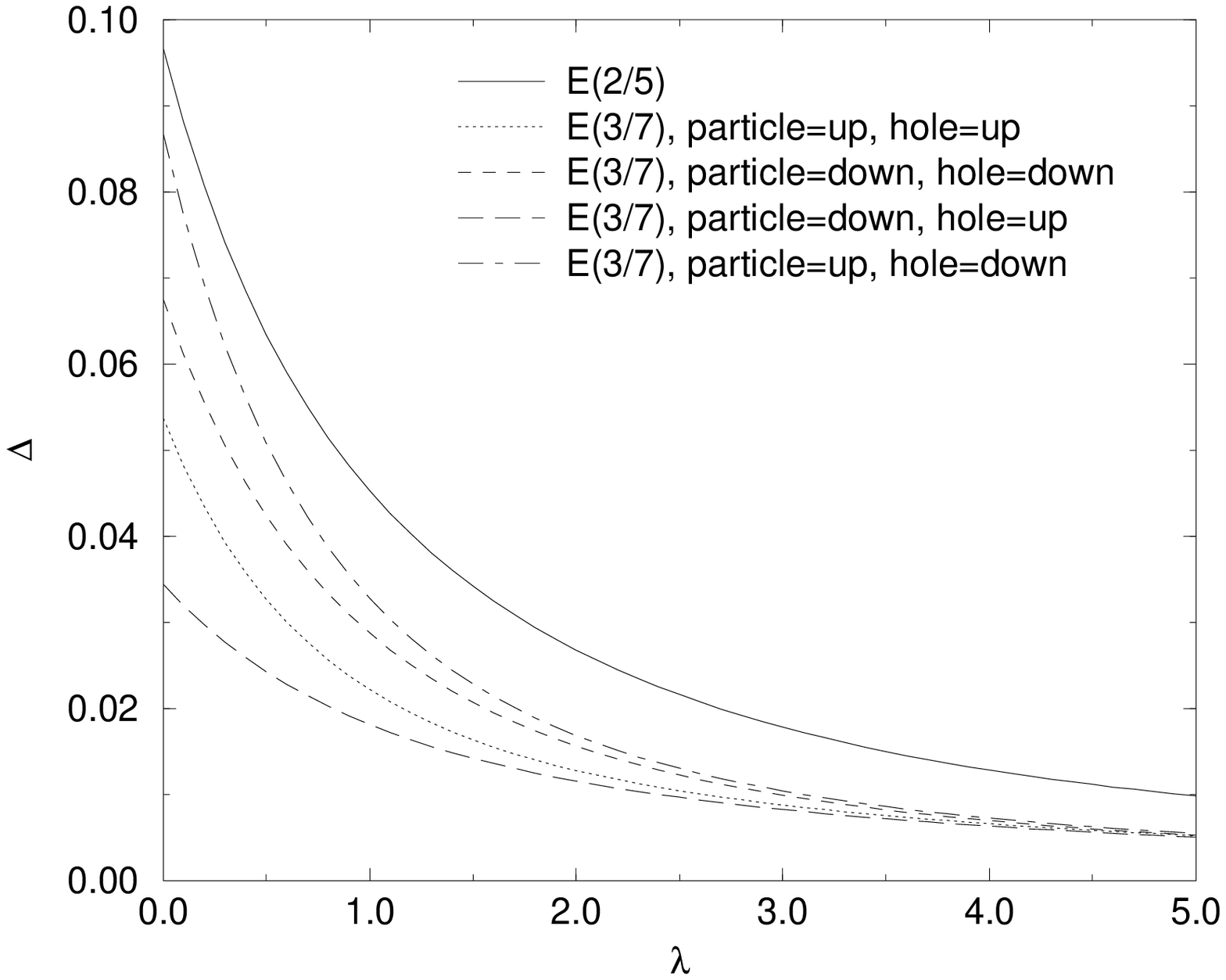}
\vskip 0.15in
\caption{Gaps to singlet ($\twof$)  and partially polarized
($\threes$) states
with constraint. All energies are in units of $e^2/\varepsilon l$.
\label{fig2}}
\end{figure}

 In summary, we have presented analytical calculations of gaps based
on our new formalism, which describes the physics directly in terms  
of
the true quasiparticles of the system, the CFs.  Central to the
analysis is the hamiltonian Eqn.~(\ref{ham}) and electronic density
Eqn.~(\ref{rhobar}) expressed in terms of CF variables. The fact that
many nonperturbative effects like charge and mass renormalization are
built in facilitates perturbative approximations.  We find agreement
with with numerical results to within 25\% (for the physical value of
$\lam$\cite{jain-th,du-numbers}) upon treating the constraint in the
simplest manner. In higher density samples, where the effective
thickness is larger, and the magnetic length smaller, $\lambda =
\Lambda /l$ will be larger. As $\lam$ increases, we expect our  
effective
theory will become more accurate, since we are using ${\bar \rho}$
matrix elements for a smaller range of $q$. We should bear in mind
that fluctuation corrections beyond HF may become important for large
$\lam$, since the higher CF-LLs come very close to the first
unoccupied one.  We also expect to run into problems as we approach
$\nu =1/2$, since constraints are known to play a crucial role in  
this
limit\cite{comment,dh,read3}.

Many other physical quantities in the gapped fractions can be  
computed
using this formalism, such as the magnetoexciton
dispersion\cite{hal-rez,GMP,he-ex,rpa,jain-ex}, which we will present
elsewhere.  Since we have a concrete microscopic description of the
CFs, we can also couple them to impurity and edge potentials.

We would like to thank S.H.Simon, S.M.Girvin, and K.Park for helpful
conversations, and especially  J.K.Jain for numerous
insightful discussions and for releasing his results prior to
publication. R.S. is grateful for support from the NSF (DMS98-00626).


\begin{thebibliography}{100}
\bibitem{fang-howard} Note that commonly used Fang-Howard
potential\cite{thick1} does not have all moments finite;
this is why we have focused on the model potential.
\end{thebibliography}
\end{document}